\documentclass[useAMS,usenatbib]{mn2e}

\usepackage{txfonts,graphicx,amssymb,rotating}

\title[Deserts and pile-ups in the distribution of exoplanets]{Deserts and pile-ups in the distribution of exoplanets due to photoevaporative disc clearing}

\author[Alexander \& Pascucci]  {R.D.Alexander$^{1,}$\thanks{email: richard.alexander@leicester.ac.uk} \& I.Pascucci$^2$ \\$^1$ Department of Physics \&
  Astronomy, University of Leicester, Leicester, LE1 7RH, UK\\$^2$ Lunar and Planetary Laboratory, The University of Arizona, Tucson, AZ 85721, USA}
  
\begin{document}
\voffset=-0.25in
% Macros
\newcommand{\Msunyr}{M$_{\odot}$yr$^{-1}$}
\newcommand{\Msun}{M$_{\odot}$}
\newcommand{\Mjup}{M$_{\mathrm {Jup}}$}

\pagerange{\pageref{firstpage}--\pageref{lastpage}} \pubyear{2012}

\date{Accepted 2012 February 24. Received 2012 February 24; in original form 2012 January 9.}

\maketitle

\label{firstpage}

\begin{abstract}
We present models of giant planet migration in evolving protoplanetary discs.  We show that disc clearing by EUV photoevaporation can have a strong effect on the distribution of giant planet semi-major axes.  During disc clearing planet migration is slowed or accelerated in the region where photoevaporation opens a gap in the disc, resulting in ``deserts'' where few giant planets are found and corresponding ``pile-ups'' at smaller and larger radii.  However, the precise locations and sizes of these features are strong functions of the efficiency of planetary accretion, and therefore also strongly dependent on planet mass.  We suggest that photoevaporative disc clearing may be responsible for the pile-up of $\sim$Jupiter-mass planets at $\sim1$AU seen in exoplanet surveys, and show that observations of the distribution of exoplanet semi-major axes can be used to test models of both planet migration and disc clearing.
\end{abstract}


%%%%%%%%%%%%%%%%%%%%%%%%%%%

\section{Introduction}
The study of planets dates back hundreds of years, but research in this area has accelerated dramatically in the decade and a half since the discovery of the first extra-solar planets \citep{mq95,mb96}.  Ground-based surveys have now discovered over 500 planets around other stars \citep[e.g.,][]{us07}, and the recent results from the {\it Kepler} satellite have pushed our census of exoplanets to over 2000 objects \citep[e.g.,][]{borucki11}.  Many of the early exoplanets were found to orbit very close to their host stars, and it was immediately recognised that these ``hot Jupiters'' could not have formed at their current locations.  Instead it is believed that planets form far from their hosts, at $\gtrsim5$--10AU, and migrate inwards due to tidal interactions with the protoplanetary disc \citep[e.g.,][]{gt80,lp86}.  The migration of low-mass planets remains controversial, but so-called Type II migration, which applies to giant planets with masses $\gtrsim 0.5$\Mjup, is now relatively well understood.  Numerical models of Type II migration show good agreement with the predictions of analytic theory \citep[e.g.,][]{tml96}, and although the origin of hot Jupiters with periods of a few days remains uncertain \citep[e.g.,][]{rice08,wl11}, the observed distribution of giant planets with $\sim$AU semi-major axes is broadly consistent with the predictions of migration models \citep[e.g.,][]{armitage07,aa09}.

How planet migration is stopped, however, remains an open question.  The migration of low-mass planets in turbulent protoplanetary discs is subject to stochastic variations \citep{np04}, and can be halted or even reversed in certain circumstances \citep{pp09}.  Type II migration, by contrast, continues as long as the disc accretes, and away from the inner disc edge it can only plausibly be halted by dispersal of the disc gas.  Disc clearing at late times is thought to be dominated by photoevaporative winds, driven by ultraviolet and/or X-ray heating by the central star \citep[e.g.,][]{acp06a,ghd09,owen10}.  These different winds disperse the gas disc and influence planet migration in quantitatively distinct ways, but observationally it is not yet clear which heating mechanism dominates \citep{pascucci11,sacco12}.  Moreover, as our census of exoplanets has grown we have discovered strong features in what were previously thought to be smooth distributions of exoplanet properties \citep[e.g.,][]{wright09}.  In this {\it Letter} we use numerical models of planet migration in evolving protoplanetary discs to investigate whether disc clearing leaves a characteristic signature on the observable properties of exoplanets.

%%%%%%%%%%%%%%%%%%%%%%%%%%%

\section{Model}
Our numerical model is essentially the same as that used in \citet[][hereafter AA09]{aa09}; here we summarise only the salient details.  We use a Monte Carlo approach, running large numbers of planet-disc models with randomly-sampled initial conditions to generate statistical distributions of planet properties.  In each individual model the protoplanetary disc evolves due to viscous transport of angular momentum and photoevaporation by the central star, while a single (giant) planet interacts tidally with the disc and undergoes Type II migration.  The equations governing the coupled evolution of the disc-planet system are \citep[e.g.,][]{lp86}
\begin{equation}\label{eq:1ddiff}
\frac{\partial \Sigma}{\partial t} = \frac{1}{R}\frac{\partial}{\partial R}\left[ 3R^{1/2} \frac{\partial}{\partial R}\left(\nu \Sigma R^{1/2}\right) - \frac{2 \Lambda \Sigma R^{3/2}}{(GM_*)^{1/2}}\right] - \dot{\Sigma}_{\mathrm {w}}(R,t) \, .
\end{equation}
Here $\Sigma(R,t)$ is the disc surface density, $t$ is time, $R$ is (cylindrical) radius, $\nu$ is the kinematic viscosity, and $M_*$ is the stellar mass.  The first term on the right-hand side describes the viscous evolution of the disc, and $\dot{\Sigma}_{\mathrm {w}}(R,t)$ is the mass-loss due to photoevaporation.  The second term describes the response of the disc to the planetary torque $\Lambda(R,a)$, which for a planet of mass $M_{\mathrm p} = q M_*$ at radius (semi-major axis) $a$ is given by \citep[e.g.,][]{armitage02}
\begin{equation}
\Lambda(R,a) = \left\{ \begin{array}{ll}
- \frac{q^2 GM_*}{2R} \left(\frac{R}{\Delta_{\mathrm p}}\right)^4 & \textrm{if } \, R < a\\
\frac{q^2 GM_*}{2R} \left(\frac{a}{\Delta_{\mathrm p}}\right)^4 & \textrm{if } \,R > a\\
\end{array}\right.
\end{equation}
where
\begin{equation}
\Delta_{\mathrm p} = \textrm{max}(H,|R-a|)
\end{equation}
and $H$ is the disc scale-height.  The planet migrates at a rate 
\begin{equation}
\frac{da}{dt} = - \left(\frac{a}{GM_*}\right)^{1/2} \left(\frac{4\pi}{M_{\mathrm p}}\right) \int R\Lambda \Sigma dR \, .
\end{equation}
We adopt an alpha-prescription for the disc viscosity
\begin{equation}
\nu(R) = \alpha \Omega H^2 \, ,
\end{equation}
where $\alpha=0.01$ is the \citet{ss73} viscosity parameter and $\Omega(R) = \sqrt{GM_*/R^3}$.  We choose a power-law form for the disc scale-height $H \propto R^{5/4}$, which gives a linear viscosity law $\nu \propto R$.  We normalize the power-law by setting the disc aspect ratio $H/R=0.0333$ at $R=1$AU.  

As in AA09, we assume that photoevaporation is driven by extreme-ultraviolet (EUV) radiation from the central star, and make use of standard prescriptions for $\dot{\Sigma}_{\mathrm {w}}(R)$.  In our standard model we adopt an ionizing flux $\Phi = 10^{42}$photons s$^{-1}$ \citep[e.g.,][]{ps09}, which results in an integrated mass-loss rate of $\dot{M}_{\mathrm w} \simeq 4 \times 10^{-10}$\Msunyr\ for $M_* = 1$\Msun.  When the inner disc is optically thick to ionizing photons (i.e., until a gap has opened in the disc), the mass-loss is concentrated around the critical radius
\begin{equation}\label{eq:R_crit}
R_{\mathrm {crit}} \simeq 0.2R_{\mathrm g} \simeq 1.8 \left(\frac{M_*}{1\mathrm M_{\odot}}\right) \mathrm{AU} \, ,
\end{equation}
where $R_{\mathrm g} = GM_*/c_s^2$ is the ``gravitational radius'' and $c_s = 10$km\,s$^{-1}$ is the sound speed of the ionized gas \citep{holl94,font04}.

The initial disc surface density profile is given by
\begin{equation}
\Sigma(R)=\frac{M_{\mathrm d}}{2\pi R_{\mathrm s} R}\exp(-R/R_{\mathrm s}) \, ,
\end{equation}
where $M_{\mathrm d}$ is the initial disc mass and the scale radius $R_{\mathrm s} = 10$AU effectively sets the viscous time-scale; for $\alpha = 0.01$, this gives $t_{\nu} = R_{\mathrm s}^2/3\nu(R_{\mathrm s}) \simeq 5\times10^4$yr.  For each model $M_{\mathrm d}$ is drawn randomly from a log-normal distribution, with mean $\log_{10}(\langle M_{\mathrm d}\rangle / \mathrm M_{\odot}) = -1.5$ and a 3-$\sigma$ spread of 0.5dex.  In the absence of a planet the median disc lifetime is $\simeq$4.5Myr, and AA09 showed that this underlying disc model is consistent with a wide range of observed properties of protoplanetary discs .  Operationally, we solve Equation \ref{eq:1ddiff} using a standard first-order explicit scheme, on a $R^{1/2}$-spaced grid covering the range $[0.04\mathrm {AU},10000\mathrm{AU}]$ (see AA09 for further details).

We model accretion across the planetary orbit in terms of the efficiency parameter $\epsilon$, which is defined as the accretion rate on to the planet as a fraction of the disc accretion rate $\dot{M}_{\mathrm {disc}}(a)$.  Numerical studies \citep[e.g.,][]{lsa99,dhk02,ld06} have shown that $\epsilon$ is a strong function of planet mass, and we adopt the fitting formula derived by \citet{va04}
\begin{equation}\label{eq:fit}
\frac{\epsilon(M_{\mathrm p})}{\epsilon_{\mathrm {max}}} =  1.67 \left(\frac{M_{\mathrm p}}{1\mathrm M_{\mathrm {Jup}}}\right)^{1/3} \exp\left(\frac{-M_{\mathrm p}}{1.5\mathrm M_{\mathrm {Jup}}}\right) + 0.04 \, .
\end{equation}
The accretion rate on to the planet $\dot{M}_{\mathrm p} = \epsilon(M_{\mathrm p}) \dot{M}_{\mathrm {disc}}$, and the accretion rate across the gap on to the inner disc is given by $\dot{M}_{\mathrm {inner}} = \dot{M}_{\mathrm p}/(1+\epsilon)$.  Our results are not especially sensitive to the precise value of $\epsilon_{\mathrm {max}}$ (we choose $\epsilon_{\mathrm {max}}=0.5$), but do depend strongly on the form of $\epsilon(M_{\mathrm p})$.  We therefore also ran models with constant values of $\epsilon(M_{\mathrm p})$, discussed below.

We ``form'' giant planets by injecting a single planet of mass $M_{\mathrm p}$ into each disc model at radius $a_{\mathrm p}$ and time $t_{\mathrm p}$.  We assume a constant formation radius $a_{\mathrm p}$, and the time of formation is assigned randomly within the range 0.25Myr $< t_{\mathrm p} < t_{\mathrm c}$ (where $t_{\mathrm c}$ is the time at which photoevaporation begins to clear the disc).  We make no attempt to model the interaction between the migrating planet and the inner disc edge; instead, we simply remove planets which migrate to $a < 0.15$AU.  As we wish to study how the properties of the surviving planets depend on their mass, we assign initial planet masses randomly in the range $0.5$\Mjup$\le M_{\mathrm p} \le 5.0$\Mjup.  This choice of a flat planet mass function serves merely to ensure we model a statistically significant number of massive planets, and is not intended to be representative of the observed exoplanet mass function (which declines steeply with increasing planet mass; e.g., \citealt{us07}).  For each choice of input parameters we ran 1000 random realisations of the model.  Typically around half of the planets migrate on to the star while the remainder survive, so each set of models results in a statistical distribution of 400--700 planets.

Our standard model considers a $M_* = 1$\Msun\ star and assumes that planets form at $a_{\mathrm p} = 5$AU.  We also ran otherwise identical sets of models with constant $\epsilon(M_{\mathrm p})/\epsilon_{\mathrm {max}} = 1.0$, 0.3, and a set of models with formation radius $a_{\mathrm p} = 10$AU.  Finally, we considered a set of models with $M_* = 0.5$\Msun.  We again form planets (arbitrarily) at $a_{\mathrm p} = 5$AU, but in this case we re-scale the disc model by setting the mean disc mass $\log_{10}(\langle M_{\mathrm d}\rangle / \mathrm M_{\odot}) = -1.8$, scale radius $R_{\mathrm s} = 5$AU, and ionizing flux $\Phi = 3\times10^{41}$photon s$^{-1}$.  There is some observational evidence for a linear scaling of disc mass with stellar mass \citep[e.g.,][]{klein03,scholz06}, but how the other model parameters scale with stellar mass in real systems is not well known \citep[e.g.,][]{aa06,kk09}. We have chosen these parameters, somewhat arbitrarily, to give a similar median disc lifetime ($\simeq 4.5$Myr) to the $M_* = 1$\Msun\ case.

%%%%%%%%%%%%%%%%%%%%%%%%%%%
 
\section{Results}\label{sec:res}
\begin{figure}
\centering
       \resizebox{\hsize}{!}{
       \includegraphics[angle=270]{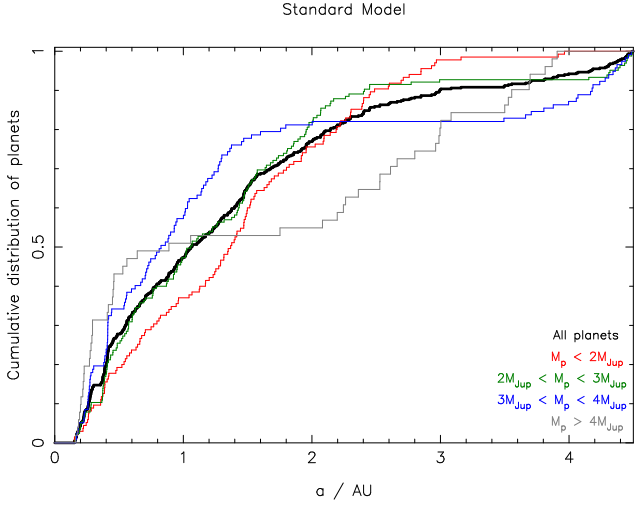}
       }
       \caption{Cumulative distribution of surviving planets in the standard model.  We plot only the planets at $a < 4.5$AU, to eliminate artefacts introduced by the (arbitrary) choice of formation radius $a_{\mathrm p} = 5$AU.  The thick black line shows the distribution for all planets, while the coloured lines denote sub-sets of this sample in different mass bins (as indicated in the legend).  More massive planets show clear ``deserts'' in their radial distributions, where few or no planets are found, close to the critical radius at which photoevaporation first opens a gap in the disc.}     
           \label{fig:standard}
\end{figure}

The radial distribution of surviving planets in the standard model is shown in Fig.\ref{fig:standard}.  If we consider the complete sample of planets we find no strong features in the distribution, and the overall shape of the distribution reflects the average migration rate \citep[e.g.,][]{armitage07}.  For a given planet mass the Type II migration rate depends primarily on the disc viscosity and surface density, so the shape of the distribution in Fig.\ref{fig:standard} primarily reflects our underlying disc model (and is consistent with AA09).  The radial distribution of planets with final radii $a < 4.5$AU in the model with $a_{\mathrm p} = 10$AU is statistically indistinguishable from the standard model: this shows that the distribution of planets in this region is determined primarily by migration, and is not very sensitive to our assumptions about where and when planets form.

When we split the planets in the standard model into mass bins, however, several strong features appear. Most prominent are clear ``deserts'', where few or no planets of a given mass are found.  The precise location of the desert is a strong function of planet mass: almost no planets are found between $a \simeq 1$--2AU for the most massive planets ($>4$\Mjup); between $a \simeq 2$--3.5AU for 3--4\Mjup\ planets; and between $a \simeq 3$--4AU for 2--3\Mjup\ planets.  There is weak evidence of a similar desert at even larger radii in the lowest mass bin ($<2$\Mjup), but this is not statistically significant (and does not persist when the formation radius $a_{\mathrm p}$ is increased).  We also see significant excesses of planets (``pile-ups'') both inside and outside these deserts, for all planet masses.

These deserts and pile-ups are the result of the interaction between the planet and the clearing gas disc, and can be readily understood.  As photoevaporation overcomes viscous accretion a gap opens in the disc at approximately $R_{\mathrm {crit}}$ (i.e., at $\simeq 1$--2AU).  
Planets interior to this gap continue migrating for a short time while the inner disc drains, and consequently few planets are found just inside $R_{\mathrm {crit}}$.  Planets at larger radii suppress disc accretion, and if this suppression is strong enough the inner disc (inside $R_{\mathrm {crit}}$) becomes optically thin, allowing photoevaporation by direct irradiation to clear the outer disc.  Planets which ``trigger'' disc clearing in this manner are subsequently unable to continue migrating, so we also see a deficit of planets just outside $R_{\mathrm {crit}}$.  When integrated over many disc-planet models this results in a desert in the radial distribution of planets close to $R_{\mathrm {crit}}$, and corresponding ``pile-ups'' at smaller and larger radii.  The final location of any given planet, however, is a strong function of the rate of accretion across the planetary orbit ($\dot{M}_{\mathrm {inner}}$).  The photoevaporative wind is only able to open a gap in the disc once its rate exceeds the local disc accretion rate.  If there is efficient accretion of gas across its orbit (i.e., $\dot{M}_{\mathrm {inner}} \sim \dot{M}_{\mathrm {disc}}$, high values of $\epsilon$) the planet has little effect on the gap-opening process, and consequently disc clearing has only a modest effect on migration.  By contrast, low planetary accretion efficiencies severely limit the disc accretion rate interior to the planet's orbit (i.e., $\dot{M}_{\mathrm {inner}} \ll \dot{M}_{\mathrm {disc}}$, small values of $\epsilon$), which accelerates the gap-opening process and has a much stronger effect on the planet's migration.  We therefore see significant pile-ups of planets of all masses (though not always at the same radius), but find that deserts close to $R_{\mathrm {crit}}$ are more pronounced for more massive planets.
 
\begin{figure}
\centering
       \resizebox{\hsize}{!}{
       \includegraphics[angle=270]{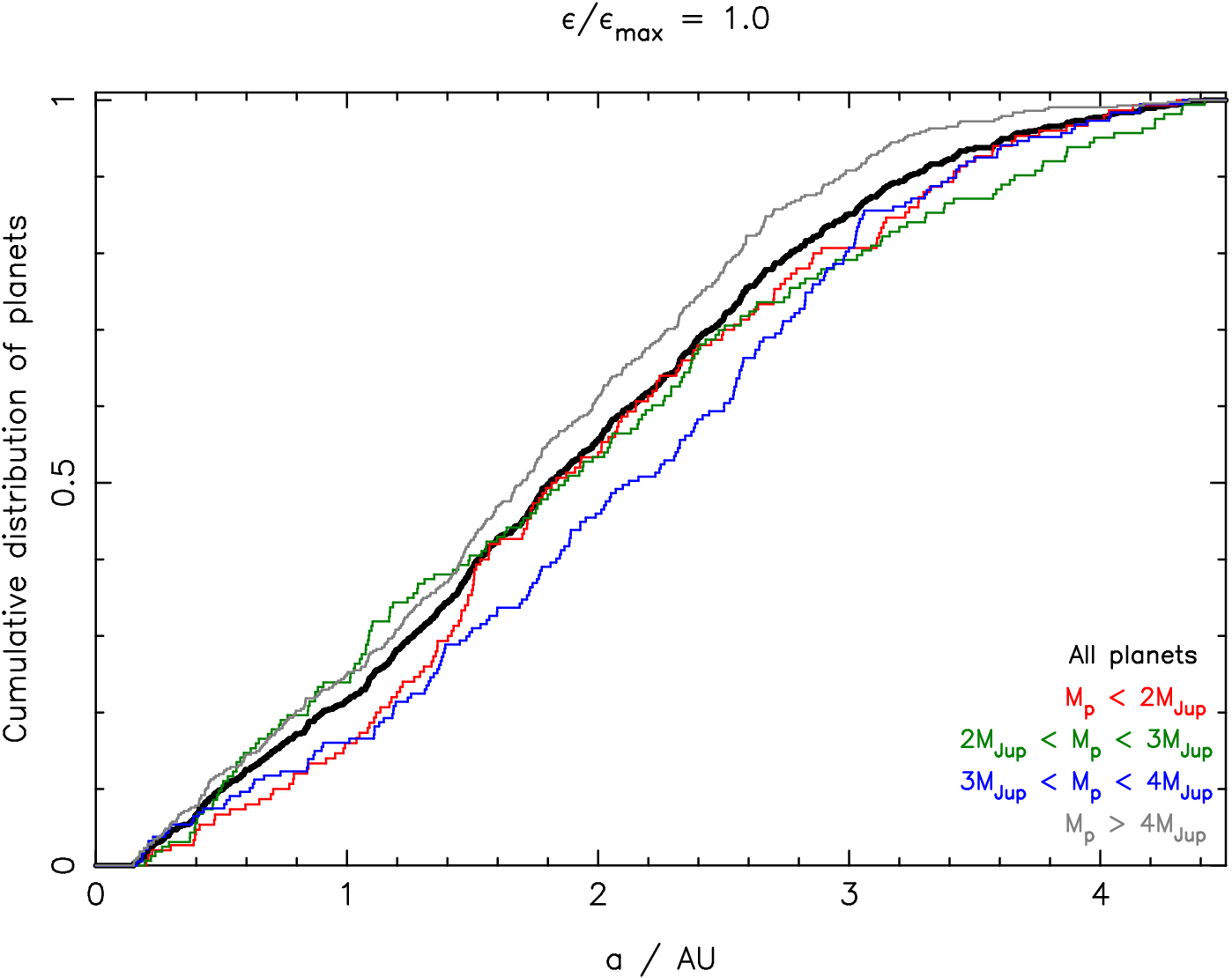}
       }
       
       \vspace*{6pt}
       
       \resizebox{\hsize}{!}{
       \includegraphics[angle=270]{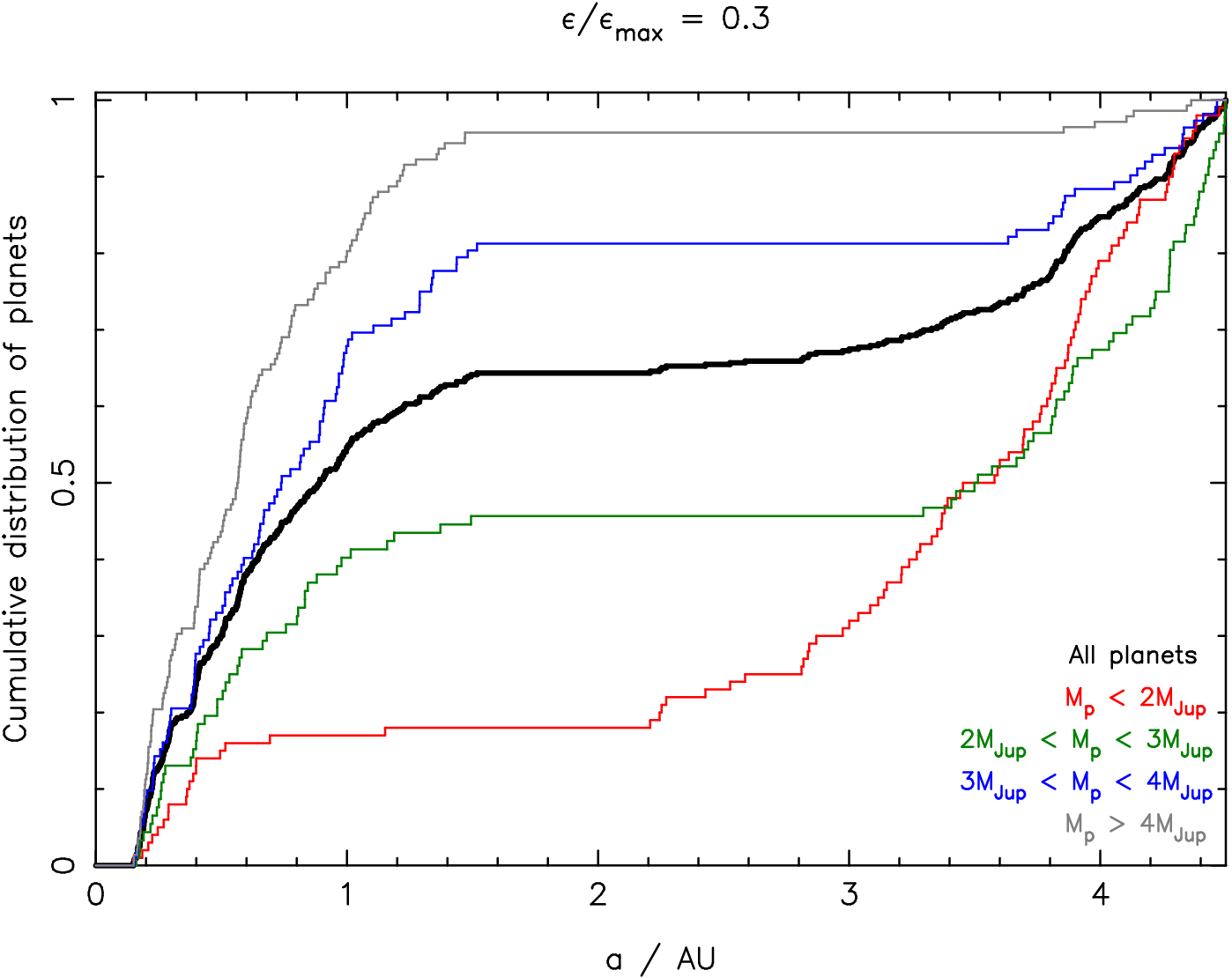}
       }
       \caption{As Fig.\ref{fig:standard}, but for models with constant $\epsilon(M_{\mathrm p})$: $\epsilon/\epsilon_{\mathrm {max}} = 1$ (top) and $\epsilon/\epsilon_{\mathrm {max}} = 0.3$ (bottom).  When gas is allowed to accrete freely across the gap (top) there are no deserts in the planet distributions, and no significant variations as a function of planet mass.  By contrast, when accretion across the gap is suppressed (bottom) we see a clear lack of planets at $a \simeq1$--3AU, and variations in the overall distribution with planet mass (because more massive planets migrate more slowly).}     
           \label{fig:epsilon}
\end{figure}
 
The sets of models with constant values of $\epsilon(M_{\mathrm p})$ allow us to investigate this behaviour in more detail (see Fig.\ref{fig:epsilon}).  For $\epsilon/\epsilon_{\mathrm {max}} = 1$ (usually consistent with $M_{\mathrm p} \simeq 0.5$\Mjup; see Equation \ref{eq:fit}) there is copious accretion across the planetary orbit for all planets, and consequently disc clearing by photoevaporation does not leave a strong imprint on the final radial distribution of planets.  Moreover, there are no significant differences in the radial distributions of planets of different masses: the distribution in each individual mass bin is statistically indistinguishable from the complete distribution.  For $\epsilon/\epsilon_{\mathrm {max}} = 0.3$ (usually consistent with $M_{\mathrm p} \simeq 2.5$\Mjup), however, we see that almost no planets with $M_{\mathrm p} > 2$\Mjup\ are found between 1.5--3AU, and very few planets with $M_{\mathrm p} < 2$\Mjup\ are found between 0.6--2AU.  There are significant pile-ups of planets inside these gaps for all planet masses.  More significant pile-ups are seen for larger planet masses: this is because more massive planets migrate more slowly \citep[e.g.,][]{sc95}, and are therefore more likely to survive at small radii (rather than migrating all the way on to the central star).  These models clearly demonstrate that the crucial parameter in determining the location of gaps in the radial distribution of planets is the accretion efficiency $\epsilon$, rather than the planet mass; the mass-dependence seen in the standard model (Fig.\ref{fig:standard}) is in fact determined primarily by the form of $\epsilon(M_{\mathrm p})$.  
 
In the models with $M_* = 0.5$\Msun\ we again find large deserts in the radial distribution of massive planets, though the effect is only apparent for planets $\gtrsim 3$\Mjup.  The inner edges of the deserts are are at smaller radii than in the $M_* = 1$\Msun\ case, but we also see that the deserts are much wider, extending from $a \simeq 1.5$--4AU for 2--3\Mjup\ planets, and from $a \simeq 1$--3.5AU for planets $>4$\Mjup.  The smaller desert radii are expected from the linear scaling of $R_{\mathrm {crit}}$ with stellar mass (Equation \ref{eq:R_crit}), and represent a possible observational diagnostic.  However, the wider deserts are primarily due to the lower disc masses in our model: massive planets open wider gaps in their parent gas discs, resulting in correspondingly wider deserts in the final distribution of planets.  Without detailed knowledge of how disc evolution scales with stellar mass it is therefore difficult to make quantitative predictions for how the giant planet population varies around stars of different mass.  Moreover, it is unclear whether this linear scaling with stellar mass can persist to lower-mass, M-type stars.  For stars of mass $M_* \sim 0.1$\Msun\ the critical radius $R_{\mathrm {crit}} \sim 0.1$--0.2AU, and is thus comparable to the radius of the inner disc edge.  In this regime the manner in which photoevaporation affects disc clearing and planet migration is far from obvious.
 
%%%%%%%%%%%%%%%%%%%%%%%%%%%

\section{Discussion}\label{sec:dis}
\begin{figure}
\centering
       \resizebox{\hsize}{!}{
       \includegraphics[angle=270]{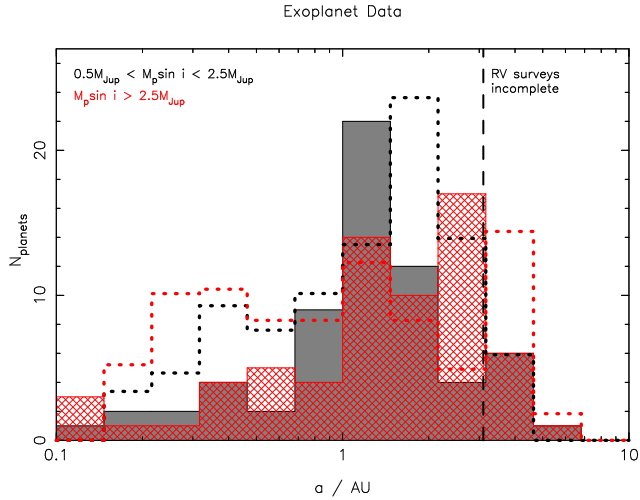}
       }
       \caption{Histogram of extra-solar planet semi-major axes.  The shaded histograms show observational data for 131 single-planet systems with reflex velocities $K > 20$\,m\,s$^{-1}$ and host star masses in the range 0.75\Msun $< M_* < $ 1.5\Msun.  The dotted histograms show the corresponding predictions of our standard model, normalised to match the numbers of planets in the observed samples.  Although the agreement is not perfect, the most prominent feature -- a significant excess of $\sim$Jupiter-mass planets with semi-major axes between 1--2AU -- is clearly seen in both the model and the data.}     
           \label{fig:data}
\end{figure}

Our major result is that protoplanetary disc clearing by photoevaporation can have a strong influence on the distribution of giant planet semi-major axes.  The general signature of this process is a gap or ``desert'' in the distribution of planets at radii close to the gap-opening radius $R_{\mathrm {crit}} \simeq 1$--2AU, with corresponding ``pile-ups'' of planets immediately inside and outside this region.  However, the precise locations and sizes of these features are strongly dependent on the efficiency of planetary accretion $\epsilon$, and consequently also on the planet mass.  There is also a weak dependence on the underlying disc model, which determines the overall rate of planet migration.  Numerical simulations have established that in the giant planet regime more massive planets accrete less efficiently, and our parametrization of $\epsilon(M_{\mathrm p})$ (Equation \ref{eq:fit}) is based on state-of-the-art numerical simulations of planetary accretion \citep[][see also Fig.1 in AA09]{lsa99,dhk02,ld06}, but it is clear from Fig.\ref{fig:epsilon} that the distribution of semi-major axes is rather sensitive to the exact form of $\epsilon(M_{\mathrm p})$.  We therefore predict that the observed distribution of exoplanet semi-major axes can be used both as a diagnostic of disc dispersal, and as a test of the physics of planetary accretion.  In particular, variations in the distribution as a function of exoplanet mass may be a useful means of determining how efficiently real planets accrete gas from their parent protoplanetary discs.  We also note that the deserts and pile-ups we predict are the result of the complicated interaction between photoevaporation, migration and accretion during disc clearing, which is usually neglected in simpler analytic calculations.  More traditional population synthesis models \citep[e.g.,][]{il08,mordasini09a} typically adopt parametrized treatments of both migration and disc evolution, which necessarily neglect these more subtle effects.  Our results suggest that a more sophisticated approach may be required if such models are to make accurate predictions of exoplanet properties.

Fig.\ref{fig:data} shows a comparison between the predictions of our standard model and current exoplanet data (taken from {\tt exoplanets.org}; \citealt{exoplanets.org}).  This compilation is not entirely free of biases, as it includes detections made using various different methods and selection criteria, but in the region of parameter space which is most relevant here ($a \gtrsim 0.1AU$; $M_{\mathrm p} \gtrsim0.5$\Mjup) the overwhelming majority of the data come from radial velocity surveys.  These data are essentially complete for reflex velocities $K > 20$\,m\,s$^{-1}$ and periods $P < 2000$\,days \citep{cumming08}, which for solar-mass stars roughly corresponds to $a < 3$AU and $M_{\mathrm p} \sin i > 1.25$\Mjup.  The observed exoplanet sample is therefore not yet ideal for rigorous statistical testing of our models (as it is incomplete for masses $\lesssim 1$\Mjup\ at $\sim$AU radii), but is appropriate for qualitative comparisons between models and data.  The major feature seen in the data in Fig.\ref{fig:data} is a significant excess of $\sim$Jupiter-mass planets with semi-major axes between 1--2AU.  This pile-up of planets is statistically significant in single-planet systems \citep[e.g.,][]{wright09}, and is not seen for higher planet masses\footnote{Strictly we are comparing the observed $M_{\mathrm p} \sin i$ values with our true masses $M_{\mathrm p}$.  However, the radii of pile-ups and deserts in our models depend primarily on $\epsilon$ rather than $M_{\mathrm p}$, so the differences between $M_{\mathrm p}$ and $M_{\mathrm p} \sin i$ are essentially degenerate with the uncertainties in the form of $\epsilon(M_{\mathrm p})$.}.  A similar feature is seen in our standard model: there is a significant pile-up of planets at 1--2AU, just interior to $R_{\mathrm {crit}}$, which is only seen for planets $\lesssim3$--4\Mjup\ (see also Fig.\ref{fig:standard}).  Although the agreement between the data and our models is not perfect it is very encouraging, especially when one considers the large number of ``un-tuned'' free parameters in the model. 

An important simplification is our choice of photoevaporation model: we assume that the photoevaporative wind is driven by ionizing extreme-ultraviolet (EUV) photons from the central star, with constant luminosity $\Phi$.  As $R_{\mathrm {crit}}$ does not depend on $\Phi$ the assumed ionizing luminosity has no significant effect on planet migration, but some recent models suggest that photoevaporative disc clearing may in fact be primarily driven by X-rays and/or non-ionizing far-ultraviolet (FUV) radiation \citep[e.g.,][]{ghd09,owen10}.  These models also predict significantly higher photoevaporation rates, up to $\dot{M}_{\mathrm w} \lesssim 10^{-8}$\Msunyr.  Observationally it is not yet clear which is the primary heating mechanism, though recent studies of diagnostic gas emission lines do not favour models where X-ray heating dominates \citep{pascucci11,sacco12}.  Our reason for not including FUV or X-ray heating is simple.  In both cases the photoevaporative flow is somewhat cooler than in the EUV case, leading to a larger critical radius $R_{\mathrm {crit}}$.  X-ray- and FUV-heated winds therefore tend to be launched from correspondingly larger radii, and their mass-loss profiles typically peak at 2--10AU or beyond.  Unfortunately, this is close to the region where we believe giant planets form by core accretion \citep[e.g.,][]{pollack96}, and consequently the effects of these winds on planet migration are degenerate with our assumptions about where and when planets form.  The general effect of X-ray and FUV photoevaporation is to create deserts in the distribution of planets between $\sim 2$--20AU, but without a fuller understanding of planet formation we cannot make detailed predictions about this process.  We note in passing, however, that large photoevaporation rates ($\dot{M}_{\mathrm w} \sim 10^{-9}$--$10^{-8}$\Msunyr) imply much larger disc surface densities at the point when disc clearing begins than are found in our EUV-only models.  In this regime planet migration is rarely suppressed, leading to much faster planet migration and making it much less likely that planets will survive interior to $R_{\mathrm {crit}}$.  Further work in this area is still needed, but our models suggest that very few giant planets should survive at small radii if photoevaporative clearing occurs at rates $\dot{M}_{\mathrm w} \gtrsim 10^{-9}$\Msunyr.

Finally, we note that photoevaporative clearing is not the only mechanism which can lead to features in the exoplanet distribution.  Various other effects, such as the location of the snow-line, formation in dead zones, or migration into planet ``traps'' \citep[e.g.,][]{masset06,mp06,il08}, can also lead to pile-ups at specific radii.  However, these features tend not to remain in fixed locations over the entire disc lifetime, and it is also not obvious if these mechanisms can reproduce the mass-dependence seen in the observations.  We predict that protoplanetary disc clearing due to photoevaporation results in a pile-up of $\sim$Jupiter-mass planets at $\sim 1$--2AU, and we suggest that this mechanism may be responsible for the similar pile-up seen in recent exoplanet surveys.  As our sample of exoplanets becomes larger and more complete, their properties will provide important observational tests of the physics of disc clearing, planet migration and planetary accretion.

%%%%%%%%%%%%%%%%%%%%%%%%%%%
\section*{Acknowledgements}
We thank Phil Armitage, Cathie Clarke \& Richard L.\,Greenberg for useful discussions and comments on the manuscript.  RDA acknowledges support from the Science \& Technology Facilities Council (STFC) through an Advanced Fellowship (ST/G00711X/1).  IP is pleased to acknowledge support from the National Science Foundation (NSF) through an Astronomy \& Astrophysics research grant (AST0908479).  This research used the ALICE High Performance Computing Facility at the University of Leicester. Some resources on ALICE form part of the DiRAC Facility jointly funded by STFC and the Large Facilities Capital Fund of BIS.

%%%%%%%%%%%%%%%%%%%%%%%%%%%

\label{lastpage}

\end{document}